\documentclass{mem}
\usepackage{natbib}\usepackage{txfonts}\usepackage{balance}
\usepackage{graphicx}
\usepackage[a4paper]{hyperref}
\idline{77:4}{000}
\begin{document}
\def\teff{$T\rm_{eff }$}
\def\kms{$\mathrm {km s}^{-1}$}

\title{ Galactic History: Formation \& Evolution }


\author{J. Bland-Hawthorn$^1$ \& K.C. Freeman$^2$}

  \offprints{J. Bland-Hawthorn}

\institute{
$^1$Anglo-Australian Observatory,
PO Box 296, Epping, NSW 2121, Australia \\
$^2$Mount Stromlo and Siding Spring Observatory,
Private Bag, Woden, ACT 2611, Australia \\
\email{jbh@aao.gov.au}
}

\authorrunning{Bland-Hawthorn \& Freeman}

\titlerunning{Galactic History}

\abstract{
We explore the motivation behind large stellar surveys in Galactic
astronomy, in particular, surveys that measure the photometric,
phase space and abundance properties of thousands or millions of
stars.  These observations are essential to unravelling the sequence
of events involved in galaxy formation and evolution, although
disentangling key signatures from the complexity continues to be
very challenging.  The new data will require major advances in our
understanding of stellar atmospheres, stellar chemistry, the dynamics
of the Galaxy and the Local Group.

}
\maketitle{}

\section{Galactic Surveys -- Why Bother?}

In this Joint Discussion, we explore the motivation behind
large surveys in Galactic astronomy. We focus our attention on
surveys that measure the photometric, phase space or abundance
properties of individual stars.  In recent years, we have seen the
release of optical and infrared all-sky photometric catalogues for
$\sim 10^8$ stars (SDSS: Gunn et al 2001; 2MASS: Cutri et al 2003),
and new astrometric catalogues from the Hipparcos satellite and the
US Naval Observatory (UCAC2: Zacharias et al 2004). The first major
kinematic stellar survey of roughly 15,000 stars was recently
completed by Nordstrom et al (2004), with two new much larger surveys
now under way (RAVE: Steinmetz et al 2006; SEGUE: Rockosi 2005).
There is on-going discussion of extending these million-star surveys
to 8m class telescopes early in the next decade (WFMOS: Colless
2005).  At that time, the European Space Agency is set to launch
the Gaia astrometric satellite with a view to establishing phase
space information for a billion stars (Perryman et al 2001; Wilkinson
et al 2005). But why go to all this trouble?

Increasingly, we are required to defend big survey machines with
rigorous and compelling science cases, and these must be argued
carefully within the context of "Big Questions" that are posted on
the web sites of major funding agencies. In fact, it is probable
that only one of these questions bears directly on galactic stellar
surveys $-$ the formation and evolution of galaxies. But here, a
good case can be made, as we discuss below.

It is no exaggeration to say that the study of galaxy formation and
evolution will dominate observational cosmology and galactic studies
for decades to come.  This is because it is difficult to define a
unique model for galaxy formation, assuming one even exists. A
theoretical astrophysicist may be content to establish the existence
of galactic "building blocks" at high redshift, and to demonstrate
that numerical simulations can explain the properties of these
objects, and the basic properties of galaxies at all epochs to the
present age. But an applied physicist may argue that this is far
from a complete picture where all of the salient microphysics is
demonstrated self-consistently.

We have long known that galaxies form over cosmic time through the
gradual build-up of dark matter and baryons within a vast hierarchy.
This process, which can be studied in the near and far field, is
likely to depend in part on the mean density field in the vicinity
of the coalescing galaxy (e.g. Maulbetsch et al 2006). There are
recent claims that large photometric surveys of galaxies argue
against strong environmental factors (Blanton et al 2006), but this
influence will be much clearer once the individual components are
properly resolved.

In the near field, the Local Group is a relatively low-density
region of the Universe; to study galaxies over a wide dynamic range
in density contrast will require that we push our stellar surveys
out to Virgo, something that is only conceivable with 30-40m class
telescopes and the James Webb Space Telescope (JWST). This is made
abundantly clear in Fig. 1 where we show the mean density of the
50 nearest galaxy groups.

\begin{figure}[t!]
\resizebox{\hsize}{!}{\includegraphics[clip=true]{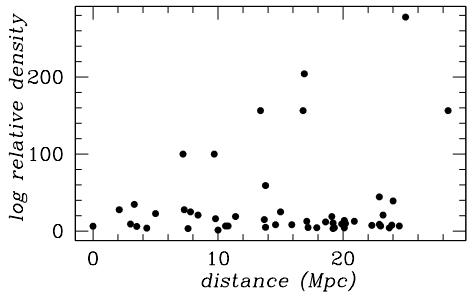}}
\caption{\footnotesize
The dependence of group density with distance from
a sample of 50 galaxy groups taken from Tully (1987).
Most groups are much like the Local Group in their average
density.
}
\label{eta}
\end{figure}

The environmental dependence is not something that can be established
easily in the far field much beyond the well known distinction of
evolved galaxies in virialized clusters and generally younger stellar
ages in the filaments between clusters (cf. Cucciati et al 2006).
But a fact well known to galactic astronomers is that essentially all
galaxies show evidence of an underlying old stellar component,
something that is difficult to resolve out in the far field.

There are numerous ways in which large-scale stellar surveys can
provide fundamental insight on the process of galaxy formation. But
first we review what has been learnt in recent years.

\section{What recent surveys have revealed}

\subsection{Resolved stellar populations}

Even in an era of Extremely Large Telescopes (ELTs), we will only
be able to resolve individual stars and stellar populations out to
Virgo (Local Volume), but this may be enough to ultimately
unravel the processes involved in galaxy formation, particularly
when supported by observations of the high-redshift universe.  Some
of the most revealing insights have come from the Local Group with
its two dominant rajahs and numerous courtiers. The Galaxy and M31
have comparable mass, and show evidence of substructure in most
stellar components (e.g. Juric et al 2005). Both have extended disks
and stellar haloes, although the former has a small bulge, while
M31 has a prominent bulge.

Some of the most remarkable observations have come from the Hubble
Space Telescope (HST) which targetted the spheroid, outer disk and
the major tidal stream in M31 (Brown et al 2006). All fields of
view were found to have an extended star formation history, with
most stars being in the age range of 4 to 10 Gyr.  The inner spheroid
is intermediate age ($6-9$ Gyr) and metal rich and slightly older
than the Stream, and may well comprise mostly stream material.  This
is in stark contrast to the mostly old Galactic spheroid.  The outer
disk in M31 is younger ($4-8$ Gyr) but still older than the thin
disk population in the Solar neighbourhood.

Resolved stellar population studies are now achieving much fainter
effective surface brightness levels than are possible with observations
of diffuse stellar light. The disks in M31 and M33 have been detected
at $\mu_V = 30$ mag arcsec$^{-2}$ and are seen to extend to more then 
10 scale lengths.  The metal-poor stellar halo in M31
is now thought to extend to at least 160 kpc (Guhathakurta et al
2005), i.e. a halo subtending 20$^\circ$ across the sky! The
halo of M33 has now been detected and found to be comparable in
metallicity to M31 and the Galaxy (McConnachie et al 2006), which
is interesting in light of the fact that it has no detectable
bulge.  At the present time, stellar haloes have not been detected
in either of the Magellanic Clouds (e.g. Gallart et al 2004).

\subsection{Thick disks \& outer disks}

It is well established that the thick disk is 10$-$12 Gyr old and
therefore provides us with an ancient snapshot of what took place
in the early universe (q.v. Reddy et al 2006).  A recent development
is that thick disks may be relatively common in disk galaxies 
(Yoachim \& Dalcanton 2005).

The character of the Galactic thick disk is now known to be distinct
from the thin disk in essentially all measurable parameters. The
dynamically hot population is characteristically older and more
metal poor than the thin disk, with an unexpectedly strong rotational
lag. There may be evidence that the thick disk has a longer scale
length (Robin et al 1996) and a distinct abundance
gradient (Brewer \& Carney 2006).

Beyond the Local Group, the closest galactic disk is NGC 300 at a
distance of 1.95$\pm$0.05 Mpc. This galaxy is a virtual twin of M33
but showing a different behaviour in the outer parts. The red stellar
disk is exponential over 10 scale lengths with no evidence
for a truncation of any kind (Bland-Hawthorn et al 2005).  In
contrast, M33 shows dramatic truncation close to the Holmberg radius
with no evidence of tidal streams (see below), quite unlike its
dominant neighbour.

Pohlen \& Trujillo (2006) have obtained SDSS photometry on 90
galaxies to find that 60\% truncate in the outer parts, 10\% remain
exponential to the limits of the data, and 30\% appear to show
flattening in the outermost parts. While these data do not reach
the same effective surface brightness as the resolved stellar
surveys, it is clear that outer disks are telling us something
important about the build-up of disk material over billions of
years. The picture is not as tidy as we used to imagine (e.g.  Fall
\& Efstathiou 1980).

\subsection{Abundance patterns and gradients}

Stellar abundances are discussed by S. Feltzing; the discussion
here is limited to a few key points.  The Hipparcos survey has
allowed a very clean kinematic separation of the thick disk from
the thin disk (Bensby et al 2005; Reddy et al 2006). This has
revealed a very clear distinction between the chemistry of the thick
and thin disk, particularly in the $\alpha$ elements.

By now, we are used to the idea of nebular abundance gradients in
spiral galaxies determined from HII regions. In contrast, stellar
populations can be aged and these appear to show an overall abundance
gradient that is flattening with time. For example, open clusters
appear to show that the Galactic abundance gradient was -0.1 dex/kpc
8 Gyr ago, softening to -0.04 dex/kpc at the present day (Daflon
\& Cunha 2004; Salaris et al 2004). This may be consistent with new
evidence of young metal-poor cepheid variables in the outer disk
(12$-$17 kpc) with enhanced [$\alpha$/Fe] and [Eu/Fe] (Yong et al
2006), suggestive of recent accretion. An interesting new development
is spatial abundance {\it maps} for a single population (e.g. Luck et al
2006).

Elemental abundances have long been argued as a key constraint on
the accretion history of satellites onto the Galaxy (Unavane, Wyse
\& Gilmore 1995).  A new development is the concept of chemically
tagging stars to a parent population from the element abundance
patterns (Freeman \& Bland-Hawthorn 2002; Bland-Hawthorn \& Freeman
2004). The basis for tagging is that stellar clusters are highly
uniform in certain chemical elements (e.g. De Silva et al 2006).
has now been demonstrated for the moving group HR 1614 originally
identified by Eggen. What is remarkable here is that HR 1614 covers
most of the sky which has led others to question the integrity of
Eggen's group. In an era of Gaia, chemical tagging will be enormously
powerful in that it will provide an independent confirmation of the
integrity of a stellar group, or allow us to disentangle cells in
phase space.

Simulations show that dwarf galaxies spiral into larger ones, where
they are torn apart to produce the star streams observed in the big
galaxies. But the patterns of heavy elements from UVES observations
at the VLT (DART: Tolstoy et al 2003) indicate that no major component
of the Galaxy could have been assembled largely by accretion of
dwarfs of the kind observed today. M31 and the Galaxy could have
formed by merging of dwarfs in the early universe; the curious thing
is that the dwarfs that were left behind to be observed as dwarfs
today have to be substantially different (Robertson et al 2005).


\subsection{Substructure \& streams}

This topic is discussed by A. Helmi so we limit our discussion here.
One of the major developments since the mid 1990s is the discovery
of the infalling Sgr dwarf (Ibata et al 1994) from radial velocity
observations at the AAT, using stellar identifications towards the
Galactic bulge identified in UK Schmidt plates. This was the first
detection of a disrupting galaxy within the orbit radius of the
Magellanic Stream which led to a resurgence of interest in Galactic
studies.  Since then, other streams have been identified by the
SDSS and 2MASS surveys although these may be associated with either
the Sgr stream (Newberg et al 2003) or the outer disk (Ibata et al
2003). 

We have become used to hearing about substructure out of the plane,
but the Hipparcos survey reveals kinematic substructure even within
the thin disk (Dehnen 1998; Fux 1997). The ages of individual clumps
are quite distinct (Famaey et al 2005) such that they are likely
to be dynamical in origin rather than due to patchy star formation.

The overlap of the Hipparcos survey with the Geneva-Copenhagen
survey allowed Navarro et al (2004) to identify a possible moving
group associated with Arcturus. They suggest an infalling group
$\sim 8$ Gyr ago.  This shows the potential for astrometric surveys
used in combination with wide-field kinematic surveys.

Interestingly, the SDSS has identified complex substructure in the
thick disk (Juric et al 2005). Tomographic slices through the galaxy
were obtained using photometric distances.  Gilmore et al (2002)
observed 2000 F/G stars with the 2dF at the AAT that were chosen
to extend to high Galactic latitude. They found that the rotation
of the thick disk is half the expected value of 180 km/s.  A possible
interpretation is that the thick disk arose from a satellite merger
10$-$12 Gyr ago.

Looking further afield, wide-field observations from the INT reveal
complex system of streams in the outer disk of M31 (Ibata et al
2005), something that is not seen in M33 (Ferguson et al 2006).
That these are discrete subcomponents has now been confirmed
kinematically using LRIS on the Keck telescope (Chapman et al 2006).

\section{Future surveys}

\subsection{The case for extending beyond the Local Group}

The observations to date provide evidence of a complex tapestry
that is only now coming into focus. We will need to study this
tapestry in far more detail before we can begin to unravel its
origins, and the processes by which it came into being.  In the
next decade, it will be possible to study the star formation histories
in hundreds of nearby galaxies, both in dense and loose groups, and
for distinct galactic subcomponents.  With these observations, we
can choose to compare individual galaxies, or compare the volume-averaged
star formation histories between groups in order to answer these
questions.

One thing is clear: we must extend these large-scale surveys beyond
the Local Group, particularly in an era of ELTs and JWST (see \S
1).  These telescopes promise diffraction limited performance in
infrared bands, which is good news for crowded fields.  But we will
also need optical bands for metallicity sensitivity in warm stars.
We will need to reach down to the main-sequence turn-off to derive
accurate ages. Remarkably, high Strehl ratios may not be required
in crowded fields (Olsen et al 2003). But just how well can one
achieve accurate photometry when using adaptive optics in long
exposures?  It is hard to guess at what the future has in store without
an answer to this troubling question.

\subsection{The origin of the thick and thin disk}

Freeman \& Bland-Hawthorn (2002) argue that establishing a
theory of galaxy formation is largely about understanding the
processes involved in forming disks in the early universe. The
ancient thick disk is a particularly attractive target because
it takes us back to an early epoch when large disks were forming
for the first time.

The origin of the thick disk remains controversial. Various scenarios
have been suggested: (i) snap-frozen relic of the old thin disk
heated by an ancient merger event; (ii) material from one or more
major merger events; (iii) dissipational collapse; (iii) the byproduct
of unbound star clusters in the early universe (Kroupa 2002).
Interestingly, Elmegreen \& Elmegreen (2006) measure the scale
heights of large star forming complexes in the Hubble Ultra Deep Field
(HUDF) and find these are broadly consistent with the
thick disk, in support of Kroupa's model.

We often think distribution functions of dominant stellar components
as being smooth for the most part. But is that really true in
practice?  In light of Juric et al (2005), important clues on disk formation 
may be revealed in the substructure.  Simulations
by Abadi et al (2003) suggest that substructure could survive
throughout the disk, presumably because much of this material spends
much of its time out of the plane. Strong scattering near resonances
may wash out some of the substructure, but equally one can imagine
fossil structures that are trapped in resonances.

In \S 2.2, we discussed new results on the disks of spirals.  Does
the extent and the specific angular momentum of the outer stellar
disk reflect certain properties of the collapsing protocloud?  If
the answer is no, then is the outer stellar disk still forming?  If
the answer is yes, what is the origin and role of the HI that extends
beyond the stellar disk in most instances?  Where stellar disks are
seen to extend far into the HI, the inferred $Q$ values are much
too high to form stars, so how is this possible?

In our view, a major shortcoming of contemporary Galactic studies
are accurate stellar ages, especially on timescales of
billions of years. One can envisage experiments to rectify this
problem for $\sim 10^5$ stars, an issue we discuss elsewhere
(Freeman \& Bland-Hawthorn 2002). The present resurgence of interest
in open clusters is partly due to the prospect of a decent age for
the stellar ensemble.  As a result,
the chemistry of stellar clusters may allow us to determine
how large-scale enrichment took place in the Galaxy over cosmic
time. Early results seem to indicate that the chemical gradient
today is flatter than it was 10 Gyr ago (\S 2.3).  An alternative
scenario to gradual disk accretion is that Galaxy-wide enrichment
events from the nucleus has gradually built up the chemical elements
in the outer disk, much like a volcano building up its ramparts
from successive events over many dynamical times.

\subsection{The origin of the bulge and halo: ancient stars}

An area of intense future interest is expected to be the spheroidal
components of M31 and the Galaxy.  Just how did these form and what is
their relation to the rest of the galaxy?  The highest resolution
simulations tell us that we can expect to find far more evidence
of substructure throughout the inner 10 kpc or so (Gao \& White
2006).  These surveys are likely to turn up an ancient stellar
populations over the entire metallicity spread, from ultra metal
poor to extremely metal rich stars.

A rare population of ancient ultra-metal-poor stars in the Galactic
halo provides critical information on the chemical yields of the
first generation of massive stars (Beers \& Christlieb 2005; Tumlinson
2006). Larger concentrations of ancient stars may be hiding in the
centres of galaxies where the mass density is high and conditions
likely first favoured star formation.  Future instruments will
search for these ancient stars, but once again, dealing with crowding
and dust obscuration will require a high spectroscopic resolution,
near-diffraction limited spectrograph working at infrared wavelengths.

A new generation of multi-object survey instruments
will be critical to unravelling some of the biggest questions in
modern cosmology today.  In addition to the fully-funded Gaia mission
in 2012, the Japanese are expected launch the JASMINE explorer in
2015 to obtain infrared astrometry on 10 million bulge stars (Gouda
et al 2005). Seidelmann \& Monet (2005) summarise a number of related
missions in the next decade. There is a great deal to be learnt on
the nature of galaxies from explorations of this kind.

\section{The importance of a dynamical framework for the Galaxy}

It is tempting to think that the billion-star Gaia survey will be
the final word on Galactic studies. However, the effective spectroscopic
magnitude limit of Gaia is only V$<$17, equivalent to a G dwarf at
a distance of 6 kpc in the absence of dust. We can go much deeper
than this with large-ground based telescopes. Such observations
will be essential if we are to extract chemical information over a
large volume of the Galaxy. Future ground-based optical and infrared
surveys will likely map the Galactic bulge and provide a self-consistent
model for the bulge and bar.  Deep pencil beam surveys towards
dust-free inner windows will establish the strength and orientation
of the spiral arms, and allow us to correct the rotation curve (and
Oort's constants) for streaming motions.

A self-consistent dynamic model of the Galaxy is essential in
understanding the smooth underlying potential.  Once we have
established the basic parameters reliably, we can revisit the
distribution of dark matter within the Solar Circle and beyond, an
essential piece of information if we are to reproduce the main
features of the Galaxy from $\Lambda$CDM simulations.  The kinematics of
the outer halo stars will ultimately constrain the shape and figure
rotation of the dark matter halo.

Binney (2004) has highlighted the need to get ready for
the impending data revolution. Now that Gaia is fully funded, the
case is even more compelling. Binney makes clear that we need a
consistent multi-component model that can be modified to give a
better fit as data become available. To complicate things further,
we suspect that these models may need to take on board the fact
that the Galaxy is not in dynamical equilibrium, and therefore to
think in the wider context of the Local Group. This may become
possible with improved 3D space motions of individual galaxies
(Brunthaler et al 2006).  Finally, Binney's vision considers only
the phase space information when in fact the chemical space is
likely to be equally unwieldy (Bland-Hawthorn \& Freeman 2004).

\section{Testing CDM in the non-linear regime}

Wide-field surveys have been highly successful in testing $\Lambda$CDM
in the linear regime to the extent that future surveys have been
proposed to explore the nature of dark energy (Seo \& Eisenstein
2003).  We can now extend this powerful methodology to resolved
studies of galaxy formation and evolution over cosmic time.

Computer simulations of how the Big Bang unfolded over 13.7 Gyr
to yield present-day galaxies can involve up to 10 billion
particles. These computations yield structures
that look a good deal like real galaxies and clusters of galaxies,
adding to the evidence that our picture for the evolution of the
universe is on the right track.  But close examination of nearby
galaxies shows discrepancies with what the simulations might lead
one to expect. For example, CDM models predict higher concentrations
of dark matter than is believed to exist in the high surface-brightness
cores of galaxies. Some of the best evidence to date has come from
attempts to construct self-consistent models of the Galactic bulge
(for a progress report, see Binney [2004]). A key conclusion is
that baryons dominate everywhere within the Solar Circle, in sharp
conflict with almost all CDM simulations.

A more sensitive test of the CDM model however requires that
we greatly increase the kinematic samples in galaxy cores beyond
the Galaxy (Gilmore et al 2006). Thus, future wide-field and
pencil-beam Galactic surveys will be necessary in order to continue
to challenge CDM calculations in the non-linear regime.

\section{A complete theory of galaxy formation?}

Inevitably, when we find ourselves swimming in a sea of complex
data, one begins to worry about fundamental limits of knowledge.
But in many applied fields, important clues to fundamental physics
often emerge from complex data, particularly when complex physical
processes are at work (cf.  beam-line experiments in particle
physics).  In fields where we are far from understanding key physical
processes, as in the study of galaxy formation and evolution, this
effort must be worth it.

Astronomers now recognize this: every large survey has unveiled
new lines of enquiry or revealed something important about our
environment.  We talk in terms of data mining, virtual observatories,
and so forth. Furthermore, the numerical simulators push down to
ever decreasing scales, and work to include new algorithms that
capture an important process. 

At what stage do we declare that galaxy formation is basically
understood? Such a declaration becomes possible when one is able
to reproduce the salient features of galaxies today, in a host of
different environments, with a theory that is firmly rooted
(presumably) in $\Lambda$CDM. This same axiomatic theory should be
able to reproduce observations of galaxies at different epochs out
to high redshift, until we reach an epoch where objects no longer
look like modern day galaxies.  The HUDF indicates that this appears
to happen at about $z\approx 2$.

{\it A moderately complete theory of galaxy formation must also
tell us whether the Galaxy is typical or unusual in any way.} It
is often stated that if the Galaxy is pathological, it is hardly
worthy of the attention we give it. But in fact, recent observations
by R.B. Tully and collaborators suggest that "Local Group" collections
of galaxies are common throughout the local universe (see Fig. 1).
Therefore, it would be a surprise to discover decades from now that
our large-scale surveys were seriously misleading us in our quest
to understand galaxy formation and evolution.

\bibliographystyle{aa}

\end{document}